\documentclass{revtex4}

\usepackage{natbib}
\usepackage[dvips]{graphicx}
\usepackage{url}

\usepackage{amssymb}

\def\figintext#1{#1}
\def\figcapatend#1{}
\def\figatend#1{}

\begin{document}




\title{Precise Orbital Tracking of an Asteroid with a Phased Array of Radio
Transponders}


\author{Bernhard W. Adams}

\affiliation{Argonne National Laboratory, 9700 S. Cass Ave.,
Argonne, IL 60439, USA}
\email{adams@aps.anl.gov}
\thanks{
This work was supported by the U.S. Department of
Energy, Office of Basic Energy Sciences under contract DE-AC02-06CH11357.
}
\date{\today}
\begin{abstract}
Deflecting an asteroid from an Earth impact trajectory requires only small
velocity changes, typically of the order of microns per second,
if done many years ahead of time. For this, a highly
precise method of determining the need, magnitude, and direction of a
deflection is required.
Although the required precision can be achieved by much less accurate
extended observations, an intrinsic resolution of $\mu $m/s permits
the live monitoring of nongravitational orbital perturbations (Yarkovsky
effect), and of a deflection effort itself.
Here, it is proposed to deploy on the asteroid's surface multiple radio units
to form a phased array capable of measuring radial velocities relative to
Earth to about 1 $\mu$m/s and ranges to 5 m.
The same technology can also be used
for scientific applications such as very-long baseline radio astronomy,
milli-Hertz gravitational wave detection, or mapping of the solar wind.
\end{abstract}
\maketitle



\section{Introduction}
Asteroid impacts on Earth are infrequent, but potentially devastating events.
One example is (99942) Apophis, a 300-meter rock, which will miss Earth in
2029 by less than the orbital height of geostationary satellites.
If, in this encounter, its center of gravity passes through the so-called
keyhole -- an imaginary,
600-meter wide region in space near Earth -- then gravitational deflection
will set it up for an impact in 2036 with an energy release similar to the
explosion of Krakatoa in 1883. Making the asteroid miss the keyhole requires
much less effort than avoiding the entire planet once the keyhole has been
traversed: Up to three years before 2029, a velocity change
$\Delta v < 1 \mu$m/s tangential to the orbit is sufficient to avoid the
keyhole \cite{sanity_check,wpdynamics}, but a $\Delta v$ of the order of
cm/s is necessary for steering away from the entire planet after a keyhole
passage. Therefore, a determination whether a
deflection will be necessary should be done well before 2029.
To obviate the need for an unnecessarily large deflection, the
orbit should be measured to a precision commensurate with the required
velocity change, i.e., 1 $\mu$m/s.
Passive radar can resolve velocities to a few cm/s and
ranges to tens of meters, and only at distances much less than the orbital
diameter of Earth and Apophis. A much higher accuracy can be achieved by use
of a radio transponder traveling with or on the asteroid.
\par
Placement and operation of a radio transponder on an asteroid is challenging
in several ways, including (i) deliverable payload, (ii) power supply,
(iii) unknown surface topography and composition, (iv) landing without bouncing
off faster than the escape velocity of a few cm/s, (v) maintaining a
directional link to Earth while the asteroid is rotating, (vi) degradation
of radio signals due to interplanetary-plasma scintillations, and (vii)
the space environment. The problem of landing would be irrelevant, and the
power supply would be simpler with a transponder aboard a
spacecraft orbiting/accompanying the asteroid, but transponders fixed to the
asteroid surface can reach a much higher accuracy (see below). Therefore,
the present proposal will concentrate on the latter approach.
\section{Design Goals}
First, to the accuracy: Doppler tracking of planetary probes has shown
performance at the stipulated level, most recently 0.5 $\mu$m/s for the
gravitational-wave experiment \cite{rs_40_3_1-9} on Cassini
currently orbiting Saturn.
Orbital refinement techniques achieve this with much less precise
individual measurements over an orbital period, but a
high-precision technology permits live
monitoring of a deflection maneuver, as well as of the Yarkovsky
\cite{sci_302_5651_1739-1742} effect (orbital perturbations due to radiation
pressure). This would also be useful for using the Yarkovsky itself
for deflection \cite{sci_296_5565_77-77}.
In addition to Doppler tracking, this proposal also aims for ranging
at 5 m resolution, corresponding to 1 $\mu$m/s over two months, about 20{\%}
of Apophis' orbital period.
These Doppler and ranging accuracies are stipulated for most of the
asteroid's orbit, except where the line of sight
passes within 100 solar radii from the sun, and radio-signal
degradation due to scintillations (see below) becomes too severe.
This makes the maximum tracking distance
1.82 astronomical units (A.U.), i.e., 2.72 $\cdot 10^8$ km.
\par
A radio transponder could be placed on the surface
or aboard a companion spacecraft orbiting or tracking the asteroid. The
latter option can be realized with proven spacecraft
technology, but it cannot reach $\mu$m/s accuracy for the asteroid's center of
mass in a single measurement: due to the irregular shape and rotation
of the asteroid, orbits are generally nonperiodic and unstable. Therefore,
long averaging is necessary without orbit correction manouvers,
but these would be required to prevent crashing
or ejection. Active locking of a companion spacecraft to
surface features of the asteroid using laser or radar ranging is also
unlikely to be accurate to the $\mu$m/s level. Therefore, instantaneous
high-precision measurements for live monitoring of deflections, or of the
Yarkovsky effect, require a transponder on the asteroid's surface.
\section{Details of the Design}
For reasonable power consumption, a directional antenna is required that
compensates for the asteroid's rotation. Using a tracking dish antenna on
the low-gravity surface seems challenging. A much better approach is found
in phased-array technology, where the phases of received and transmitted
signals in a multitude of dipole antennae are controlled electronically without
any mechanical motion for directionality through constructive interference.
The phased array discussed here \cite{apo_prop} consists
of 475 radio units grouped in 19 nodes of 25 each (Fig. \ref{f_nodes}).
These numbers are somewhat arbitrary, and are meant only as an example.
\figintext{
\begin{figure}
\includegraphics[scale=0.5]{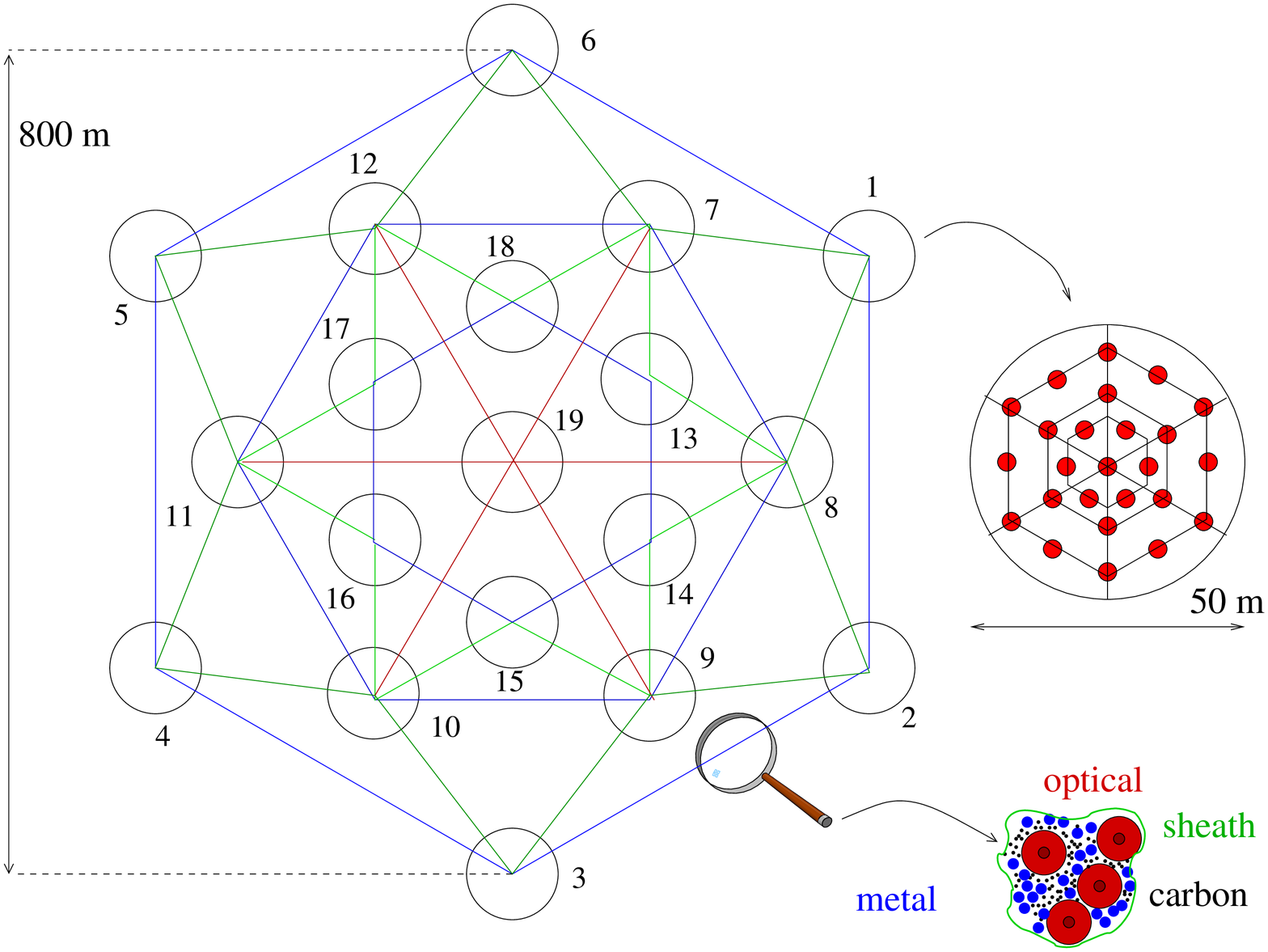}
\caption{\label{f_nodes}Left: The large web connecting 19 nodes by
optical fibers and thin electrical wires. Right: One node containing
25 transponders, connected to each other by optical fibers and thin metal wires
attached to a thin plastic foil. As shown, the total length of wire connecting
the nodes is 11.4 km. Lower right: cross section of a wire with carbon fibers,
optical fibers, and metal filaments.}
\end{figure}
}
The nodes and the radio units within each node are linked together by a net of
thin threads containing carbon fibers for mechanical strength, glass
fibers for the distribution of a precise timing signal (for phasing)
and digital communication, and metal wires for
electric-power sharing. Each radio unit has 24 antennae for
a total of 11400 for each of the frequencies used (see below). Each
node contains a highly stable oscillator, one of which generates a
reference frequency that is distributed over the glass-fiber
network. Shortly before arrival, the transfer spacecraft
aims for a collision with the asteroid at a velocity of about 0.1 m/s
(roughly the escape velocity from the asteroid's surface).
As shown in Fig.\ \ref{f_bus}, it then sequentially ejects 18 containers
\figintext{
\begin{figure}
\includegraphics[scale=0.45]{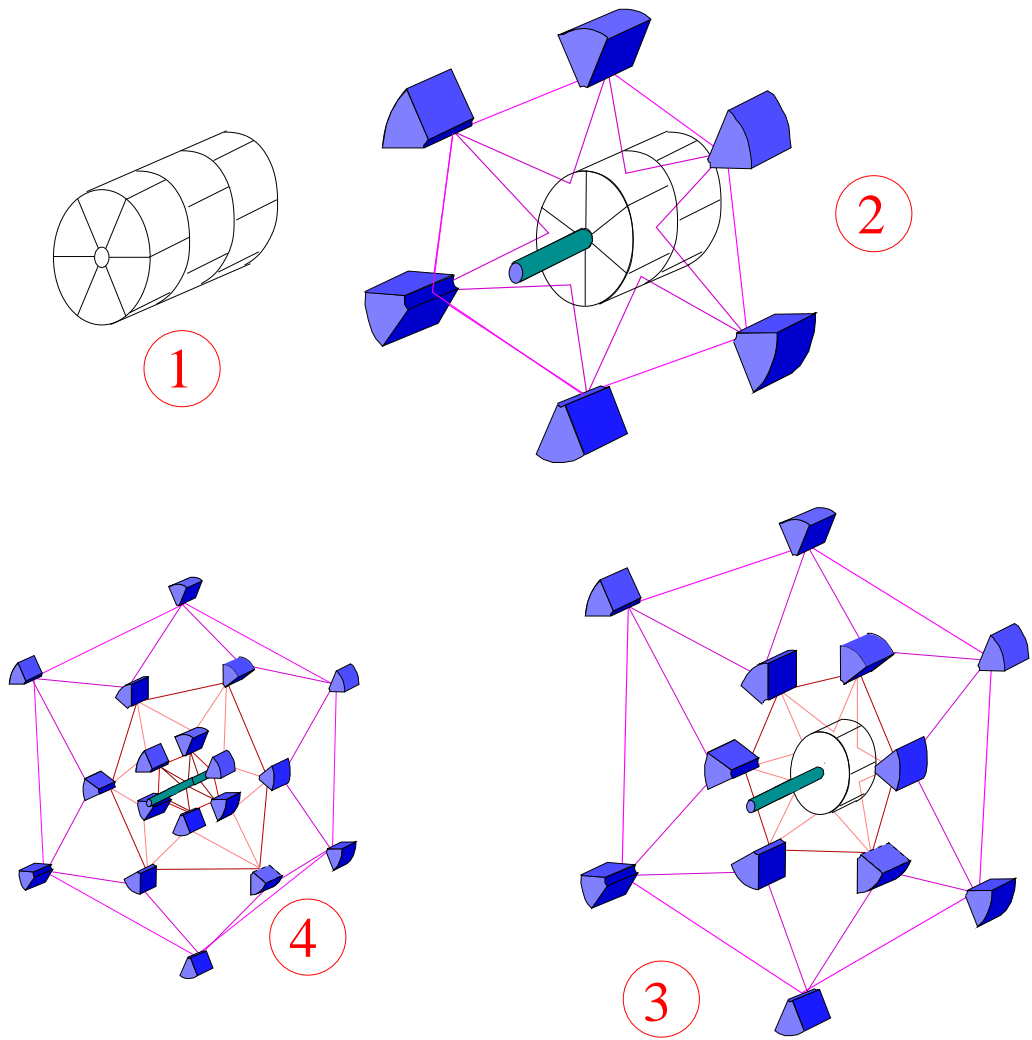}
\caption{\label{f_bus}Deployment of the
coarse web. A space vehicle (1) carries the array to the asteroid. It is
comprised of three sections, each containing six
slices. These are ejected in sequence (2-3-4) to deploy the web.
After ejection of all slices, the entire web is made to rotate slowly to
keep it stretched out (see text). Finally, each of the slices deploys a
sub-web, as shown in fig.\ \ref{f_unfold}.}
\end{figure}
}
for the nodes, which are connected by carbon/metal/glass-fiber threads.
To keep the fully deployed net spread out over about 800 meters, it is then
spun up by gas
jets to about one revolution per hour. Next, each container (including one
in the remainder of the spacecraft)
releases a thin plastic foil with attached radio units and inflates it
to 50 m diameter by injecting gas into thin tubes (Fig.\ \ref{f_unfold}).
\figintext{
\begin{figure}
\includegraphics[scale=0.55]{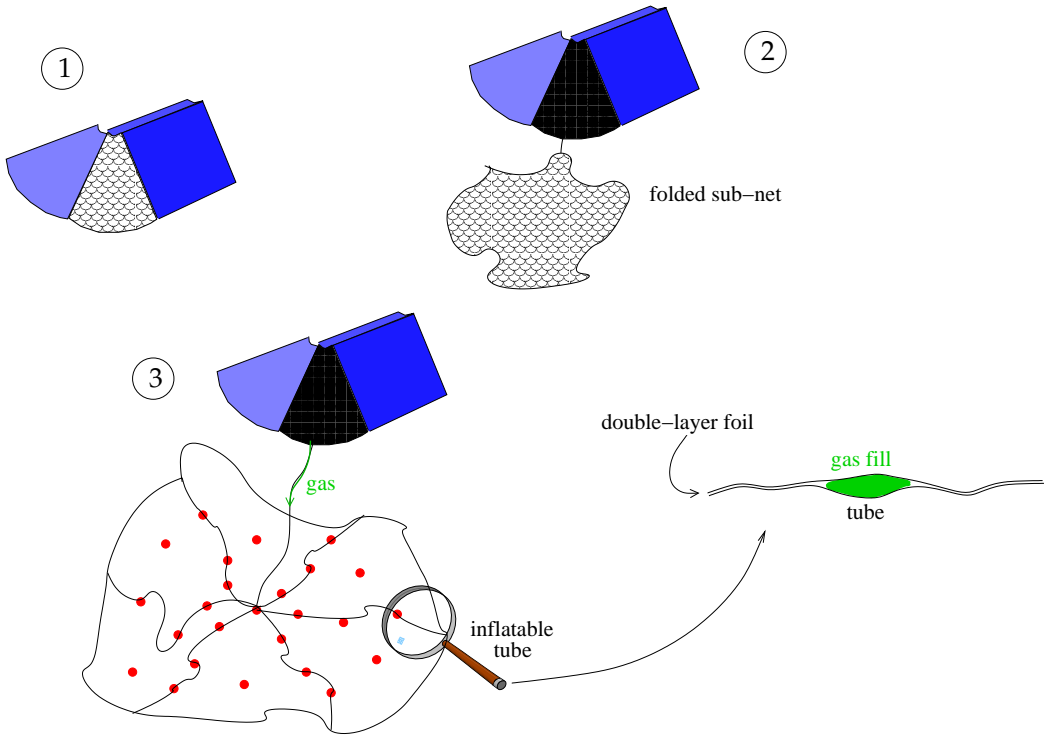}
\caption{\label{f_unfold}Ejection and unfolding of the sub-webs:
1: container opens, 2: folded plastic foil is ejected, 3: foil unfolds as gas
is blown into tubes formed between two layers of the foil.}
\end{figure}
}
Continuing on its collision course, the assembly wraps itself around the
asteroid. After several slow-motion bounces that are restrained by the net,
everything settles down in fixed random locations. Some damage is tolerable
as it only leads to a slightly degraded performance (see below).
\par
Next, Earth sends a radio signal sufficiently strong for reception without
phasing (see below). As the asteroid rotates, the array compares the phases
of this signal in each antenna to the local reference, and thus acquires
the information necessary for phased-array operation at each of the
frequencies used. This information is continually refined as the asteroid
and Earth progress along their orbits.
\par
The Doppler-tracking resolution is
degraded by scintillations due to fluctuations of the radio refractive
index n of the turbulent interplanetary plasma (solar wind).
Extrapolating spacecraft-communication data \cite{apj_219_1_727-739} for
a frequency of $\nu = 2.3$ GHz from 20 to 100 solar radii,
and applying scaling \cite{ga02__1278-1287} with $\nu^{-1.2}$, the
spectral broadening due to scintillations of an 8-GHz signal at 100
solar radii is about 0.01 Hz.
Ways to mitigate them are discussed in the literature on spacecraft-based
gravitational-wave experiments \cite{apj_599_1_806-813,lrr_9_1_1-56},
using the dispersion $n-1 \sim n^{-2}$ of $n$ with multiple frequencies $\nu$
to measure, and then correct for the fluctuations.
Here, frequencies of about 8 GHz and 16 GHz will be
used (slightly offset in up- and downlink).
\par
Doppler shifts are detected through phase shifts between ground and remote
clocks, and a radial-velocity resolution of $\Delta v_r = 1\mu$m/s requires the
reference clocks to be stable to $\Delta v_r/c = 3 \cdot 10^{-15}$,
where $c$ is the speed of light. With a radio frequency of
$\nu = 8\mbox{ GHz}$ and a phase sensitivity of $s=2\pi / 100$ (3.5 degrees),
this stability must be maintained over the time
$\tau = (c/\Delta v_r) s / (2\pi \nu) = 375$ s.
This exceeds the state of the art of 1 part in $10^{-13}$ over 1000 s in
spaceborne ultra-stable oscillators (USOs), but can be achieved with
a coherent link \cite{sci_183_4131_1297-1301,apj_210_1_568-574},
where the uplink signal is used
to phase-lock the space-borne oscillator from which the return signal is
derived. Recently, the modified technique of noncoherent Doppler tracking
\cite{aas_57_2_540-553} was used, where the spacecraft
has a freely-running oscillator and transmits its phase slippage relative
to the signal received from Earth through a digital communication channel.
This latter technique is proposed here \cite{apo_prop} to make it easier to
accommodate radio signals at two frequencies that
fluctuate differently as they traverse the solar wind. With the above
oscillator stability of $10^{-13}$, the 1-$\sigma$ walk-off of an 8-GHz
signal is
$2\pi$ in 1250 s. Thus, to resolve the phase to $s = 2 \pi /100$, about
8 bits of digital data need to be transmitted every 1250 s for 1-$\sigma$
certainty, and 10 bits for 4-$\sigma$.
\par
For ranging, Earth and the array transmit radio pulses to each other with a
pseudo-random number (PRN) phase modulation similarly to GPS, and
cross-correlate the received signals with the local ones.
The array transmits the correlation result through the digital
communication channel. With Gaussian Minimum-Shift Keying (GMSK) and a
bit-time-bandwidth product of BT = 0.3 (standard in cell
phones\cite{gsm-doc}), the signal bandwidth for a 5-meter resolution
is about 20 MHz \cite{apo_prop}. Here, $\nu=2.25$ GHz is chosen for the
uplink, and $\nu = 2.1$ GHz for downlink ranging.
The pulse duration must cover the initial ranging uncertainty
of about 1 km, that is 3 $\mu$s, but shorter PRN sequences may be used
once the range is known better.
The radio system thus uses four frequencies in each direction:
two for Doppler tracking, one for ranging, and one for digital communication.
Up- and downlink frequencies are slightly offset from each other
to isolate the receivers at each end from the transmitters. Each of the
up/down frequency pairs has 11400 antennae in the array.
\par
The transmitter powers of the array and the ground station are determined
by their mutual distance $d$, the ground and array antenna gains $G_g$ and
$G_a$, and the receiver sensitivities.
On average, half of the 11400 antennae in the array are
in view of the Earth at a given time, and their random orientations
introduce
\cite{apo_prop} another
factor of $1/4$, so $G_a=10 \log(11400/8)=31.5 \mbox{ dB}$ for all wavelengths.
Signal diminuition with distance is described by the free-space loss
$L=(4\pi d/l)^2$, i.e., the reciprocal solid angle subtended by a dipole
antenna at one end of the radio link, as seen from the other.
For a maximum operating distance of 1.82 AU at a frequency of 8 GHz,
${\cal L} = 10 \log_{10} L = -279 \mbox{ dB}$. Finally, the receiver on
Earth is assumed to use a 70-meter dish antenna (gain $G_g=65 \mbox{ dBi}$
at 8 GHz) of the deep-space network \cite{dsn_status}.
The sum $G_a+G_g+{\cal L}$ gives a total
transmission loss of -182.5 dB for all radio frequencies (${\cal L}$ and
$G_g$ go opposite in frequency, $G_a$ is frequency-independent).
The receiver system noise temperature of the ground station is estimated
\cite{ieee_ae_5_5_73-76} at 17 K, corresponding to a spectral power density
of -186.3 dBm/Hz.
For downlink Doppler tracking, this noise power needs to be matched within a
scintillation-broadened bandwidth of 0.01 Hz. With a 20-dB noise margin and a
transmission loss of -182.5 dB, the array thus needs to output -3.8 dBm
(0.4 mW) for each of the two radio frequencies. The design should foresee a
slightly higher power to allow for a reduction in array gain due to a
potential loss of radio units.
For downlink ranging at 20-MHz bandwidth and 0-dB noise margin, the transmitter
pulse power is 69 dBm (8.3 kW), to be shared among -- on average --
5472 antennae. Each antenna driver thus outputs a peak power of
about 1.5 W, i.e., about as much as a cell phone.
With 3-$\mu$s-long ranging pulses transmitted once in 30 ms, the average
power for ranging is 800mW for the array, 3.4 mW per radio unit (assuming
235 of 475 being in view of Earth and active), and 140 $\mu$W per antenna.
The lack of noise margin is compensated by repeated ranging pulses:
10000 pulses for a 20dB margin taking 300 s define the time for one
ranging measurement. Once the range is known to tens of meters, shorter
PRNs can be used for lower average power or improved noise margin.
Loss of some radio units does not disable ranging, but
leads to an increase in the measurement time.
\par
For the uplink, a higher transmitter power is available, but the receiver
noise temperature is much higher (assumed here 1000 K , which is typical
for cell phones).
Thus, the ground station has to transmit at 60
times the downlink power, i.e., bursts of about 0.5 MW for
ranging and a continuous signal of 24 mW for Doppler tracking. The
initial-phasing signal has to be received without array
gain ($G_a=0 \mbox{ dBi}$), and it needs to have a bandwidth given by the
Doppler-shift uncertainty $\Delta \nu$ due to an initial velocity uncertainty
$\Delta v$. For $\Delta v=1 m/s$ and $\nu = 8$ GHz,
$\Delta \nu=27 Hz$, and thus the ground station has to transmit at 1.6 kW for
a noise margin of 20 dB.
\par
Power is generated by radiation-tolerant, thin-film, amorphous-silicon
solar cells deposited on the plastic foils of the nodes (see
Fig.\ \ref{f_node}). This technology achieves a power of 2 kW per kg
of solar-cell/substrate material \cite{unisolar} for operation at 1
A.U.\ distance
from the sun.
Outside of the solar-cell areas, the foils have a broadband dielectric
mirror coating to reflect visible and near-infrared sunlight, but let
far-infrared escape. This reduces the temperature under the coating to well
below the 321 K in black-body radiation equilibrium at 1 A.U.\ from the sun,
so the radio unit can shed excess heat.
\figintext{
\begin{figure}
\includegraphics[scale=0.54]{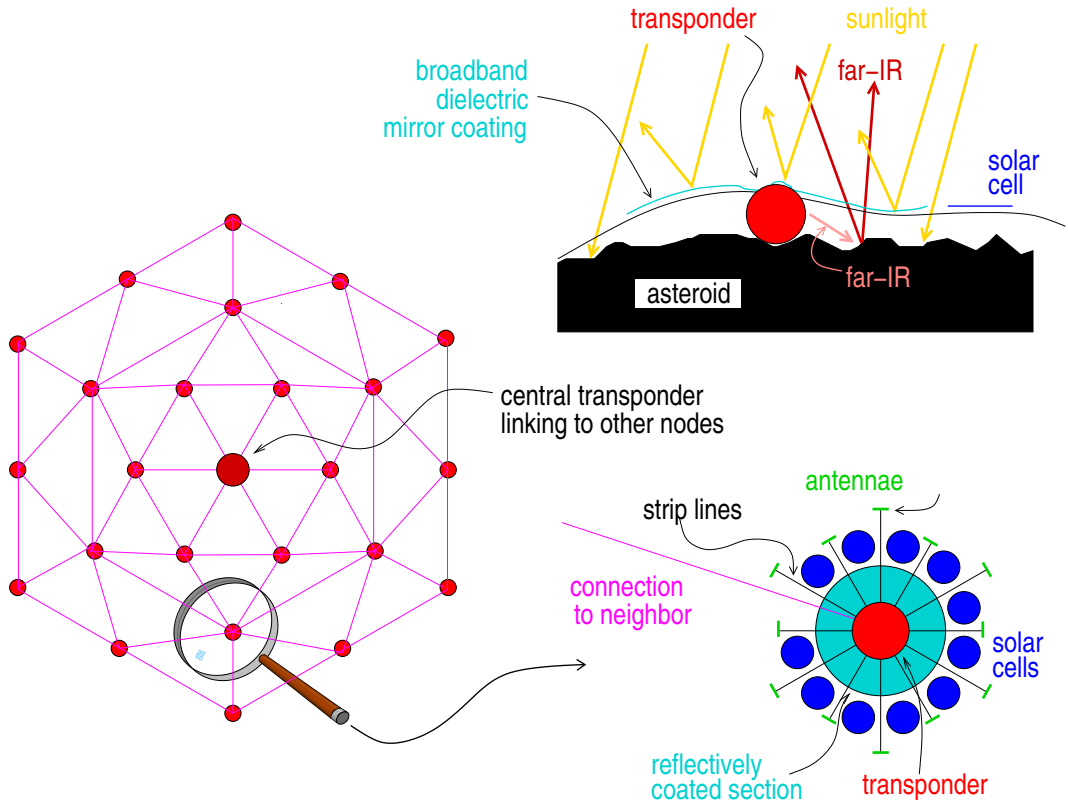}
\caption{\label{f_node}Left: The plastic foil of one node with
25 transponders sitting on it. Lower right: Closeup of the foil surrounding
one transponder with microwave strip lines going out to the antennae, and a
section of the plastic foil coated with a dielectric-mirror multilayer that
reflects visible and near-infrared sunlight, but lets far infrared through.
Upper right:
Side view of the same. Sunlight is reflected off the coating, but far
infrared heat radiation can escape into space. The transponder can thus
radiate heat off to the ground surrounding it.}
\end{figure}
}
At any given time, about half the radio units are in the sunlight and have
to power themselves and the other half through the metal wires
between and within the nodes, which are switched into rapidly varying
configurations of closed circuit loops containing power sources and sinks.
At an efficiency of 10 \%, the radio transmitters will consume an average
power of 50 mW (10 times (3.4 mW for ranging, 2 $\mu$W for Doppler, 1.6 mW for
digital communincation), 50 mW are needed for the onboard digital electronics,
and 900 mW for cooling radio units in the sunlight, so a radio unit in the
sunlight needs to harvest 1 W for itself.
Radio units in the shade need 100 mW for the electronics, 100 mW power
reserve for heating, and 100 mW for internal power conversion losses.
As estimated below, a sunlit radio unit then needs to harvest
another 2.4 W to power up to two units in the shade.
With wires made of a high-strength aluminum alloy with a specific resistance
of $2.9 \cdot 10^{-8}$ $\Omega$m \cite{apo_prop} and a cross section of
$2.0\cdot 10^{-8}$ m$^2$ (250 wires of 10 $\mu$m diameter, each),
the wires have a resistance of 1.48 $\Omega$/m.
For example, in a 1250-meter-long current loop with two sunlit nodes powering
four in the shade (each with 25 radio units), a current
of 100 mA needs to run 50 \% of the time to supply 30W at a voltage drop of
600 V (150 V per node). There is an additional voltage drop of 185 V in
the wires. Therefore, each of the two sunlit nodes need to generate
a voltage of 393 V to power the loop. At 66 \% power-conversion efficiency
each sunlit node has to harvest about 60 W for those in the shade,
i.e., 2.4 W per radio unit. For more detail, see ref.\ \cite{apo_prop}.
\par
To determine the payload to be delivered to the asteroid, the following
masses are estimated \cite{apo_prop}: 47.5 kg for the 475 radio,
5 kg for the plastic foil of the 19 nodes (excluding the solar cells),
0.81 kg of solar cells (475*3.4 W / (2 kW/kg)),
1.1 kg for 12 km of carbon/glass fiber/metal wires, 6 kg for the
power-conversion electronics, master clocks and other infrastructure
in the nodes, and 9.5 kg for the
containers in which the nodes are stowed until they are deployed
(Fig.\ \ref{f_bus}). This brings the total array mass to about 70 kg.
Another 80 kg can be estimated for the delivery spacecraft
(structure, engines, tanks, etc.) for a mass $m_s  = 150 $ kg to be delivered
to the asteroid. An optimal transfer trajectory from low-earth orbit (LEO)
to Apophis requires a total velocity change of $\Delta v = 7.71$ km/s
\cite{apo_prop}. Applying the Tsiolkovsky rocket equation
$m_t = m_s \exp (\Delta v / v_p)$ for the total mass $m_t$ at LEO, where
$v_p$ is the rocket exhaust velocity (3.07 km/s for hydrazine thrusters),
the mass to launch to LEO is about 1900 kg, which is well within the
capabilities of commercial satellite launchers.
\section{Other Possible Applications}
Space-based phased arrays -- based on asteroids, or floating freely --
could also be used for other scientific
applications, especially so if multiple ones are used. These applications make
use of the capability of a phased array to transmit and receive simultaneously
in different directions.
One example is very-long-baseline (VLBI) radio astronomy
\cite{apo_prop} with baselines the size of Earth's orbit. The arrays would
maintain radio links with each other to distribute a master clock signal,
and would simultaneously receive directionally from the radio source of
interest. As the asteroids progress along their orbits, different Fourier
components of the source's angular distribution on the sky can be resolved
to assemble an extremely high-resolution image, corresponding to the
spacing between the asteroid-based receivers.
Another possible application is in gravitational-wave detection at milli-Hertz
frequencies, as is currently being done with the Cassini
spacecraft \cite{rs_40_3_1-9}. Because this application requires a very
high velocity accuracy, the arrays have to be based on asteroids to minimize
their susceptibility to orbital perturbations due to the solar wind, the
Yarkovsky effect, etc.
With phased arrays based on several asteroids, multiple
coherent links can be maintained similtaneously in different directions
between asteroids to precisely locate a source \cite{GW_pa}.
Yet another possible application would be in the mapping of the solar
wind by measuring the scintillations in radio links between several arrays.
The transmitter power for the latter two applications may be estimated as
before for the downlink in Doppler tracking, with the bandwidth again given
by scintillations. However, the receiver temperature is now 1000K (instead
of 17 K), and the receiver antenna gain is 31.5 dBi (instead of 65 dBi).
This requires an increase of the transmitter power by a factor of
$1.1\cdot 10^5$ over the value of 0.4 mW estimated before. Thus, for each link
the array must transmit at 45 W, and each of the about 200 to 250 radio units
in view of the respective receiver must transmit about 200 mW.
These scientific applications are not subject to the time constraints of a
mission to track an asteroid before an impact or a keyhole passage.
Therefore, lengthier but more energy-efficient transfer trajectories
\cite{bams_43__43-73} are admissible, which will drastically reduce the
mass to launch to LEO, and thus the mission cost.
\section{Summary}
In summary, a way is proposed to precisely track the orbit of an asteroid
by deploying a radio transponder on its surface, which establishes a
two-frequency noncoherent link for Doppler tracking at 1 $\mu$m/s accuracy,
returns broadband-modulated pulses for ranging to 5 m, and communicates
digital data representing correlations between uplink and downlink
Doppler and ranging signals. The transponder is a phased
array of 11400 antennae for each of the four frequencies used, driven by
475 radio units. These are linked with each other by a net that serves the
multiple purposes of enabling the landing maneuver in the low gravity of the
asteroid, phasing the array, and sharing solar-generated power.
Due to a high degree of redundancy, the array can tolerate some damage, which
will only lead to a slight reduction in antenna gain. The same phased-array
technology can also be used for very-long-baseline radio astronomy or for
gravitational-wave detection.



\bibliographystyle{apsrev}
\bibliography{apophis,phlit}

\end{document}